# Inorganic Scintillators for Future HEP Experiments

Chen Hu, Liyuan Zhang and Ren-Yuan Zhu*[a]
[a] 256-48, HEP, California Institute of Technology, Pasadena, CA 91125, USA

## ABSTRACT

Future HEP experiments at the energy and intensity frontiers present stringent challenges to inorganic scintillators in radiation tolerance, ultrafast time response and cost. This paper reports recent progress in radiation hard, ultrafast, and cost-effective inorganic scintillators for future HEP experiments. Examples are LYSO crystals for a precision time of flight detector, LuAG ceramics for an ultracompact, radiation hard shashlik sampling calorimeter, BaF$_2$:Y crystals for an ultrafast calorimeter, and cost-effective scintillators for a homogeneous hadron calorimeter. Applications for Gigahertz hard X-ray imaging will also be discussed.

## INTRODUCTION

Historically, inorganic scintillators are widely used in the high energy physics (HEP) experiments to construct electromagnetic calorimeter, providing the best possible energy resolution and position resolution, good electron and photon identification and reconstruction efficiency. A homogeneous hadron calorimeter (HHCAL) would also provide excellent jet mass resolution by dual readout of either scintillation and Cerenkov light or scintillation light in fast and slow gate. The recent DOE report on HEP detector R&D needs [1] points out the following priority research directions which are related to the inorganic scintillators. Fast and radiation hard inorganic scintillators are needed to face challenges of the unprecedented harsh radiation environment expected by future HEP experiments at the energy frontier, such as the High Luminosity Large Hadron Collider (HL-LHC) and the proposed Future Hadron Circular Collider (FCC-hh), where up to 500 Grad and $5\times10^{18}$ n$_{eq}$/cm$^2$ of one MeV equivalent neutron fluence are expected by the forward calorimeter [2]. Ultrafast inorganic scintillators are needed for future HEP experiments at the intensity frontier to mitigate high event rate and pileup [3]. Cost-effective inorganic scintillators are needed for crystal ECAL and the proposed HHCAL concept. Table 1 lists optical and scintillation properties of fast and ultrafast scintillators.

**Table 1** Optical and scintillation properties of fast and ultrafast inorganic scintillators

| | BaF$_2$ | BaF$_2$:Y | ZnO:Ga | YAP:Yb | YAG:Yb | β-Ga$_2$O$_3$ | LYSO:Ce | LuAG:Ce | YAP:Ce | GAGG:Ce | LuYAP:Ce | YSO:Ce |
|---|---|---|---|---|---|---|---|---|---|---|---|---|
| Density (g/cm$^3$) | 4.89 | 4.89 | 5.67 | 5.35 | 4.56 | 5.94 | 7.4 | 6.76 | 5.35 | 6.5 | 7.2[f] | 4.44 |
| Melting points (°C) | 1280 | 1280 | 1975 | 1870 | 1940 | 1725 | 2050 | 2060 | 1870 | 1850 | 1930 | 2070 |
| X$_0$ (cm) | 2.03 | 2.03 | 2.51 | 2.77 | 3.53 | 2.51 | 1.14 | 1.45 | 2.77 | 1.63 | 1.37 | 3.10 |
| R$_M$ (cm) | 3.1 | 3.1 | 2.28 | 2.4 | 2.76 | 2.20 | 2.07 | 2.15 | 2.4 | 2.20 | 2.01 | 2.93 |
| λ$_I$ (cm) | 30.7 | 30.7 | 22.2 | 22.4 | 25.2 | 20.9 | 20.9 | 20.6 | 22.4 | 21.5 | 19.5 | 27.8 |
| Z$_{eff}$ | 51.6 | 51.6 | 27.7 | 31.9 | 30 | 28.1 | 64.8 | 60.3 | 31.9 | 51.8 | 58.6 | 33.3 |
| dE/dX (MeV/cm) | 6.52 | 6.52 | 8.42 | 8.05 | 7.01 | 8.82 | 9.55 | 9.22 | 8.05 | 8.96 | 9.82 | 6.57 |
| λ$_{peak}$[a] (nm) | 300 / 220 | 300 / 220 | 380 | 350 | 350 | 380 | 420 | 520 | 370 | 540 | 385 | 420 |
| Refractive Index[b] | 1.50 | 1.50 | 2.1 | 1.96 | 1.87 | 1.97 | 1.82 | 1.84 | 1.96 | 1.92 | 1.94 | 1.78 |
| Normalized Light Yield[a,c] | 42 / 4.8 | 1.7 / 4.8 | 6.6[d] | 0.19[d] | 0.36[d] | 6.5 / 0.5 | 100 | 35[e] / 48[e] | 9 / 32 | 115 | 16 / 15 | 80 |
| Total Light yield (ph/MeV) | 13,000 | 2,000 | 2,000[d] | 57[d] | 110[d] | 2,100 | 30,000 | 25,000[e] | 12,000 | 34,400 | 10,000 | 24,000 |
| Decay time[a] (ns) | 600 / 0.5 | 600 / 0.5 | <1 | 1.5 | 4 | 148 / 6 | 40 | 820 / 50 | 191 / 25 | 53 | 1485 / 36 | 75 |
| LY in 1$^{st}$ ns (photons/MeV) | 1200 | 1200 | 610[d] | 28[d] | 24[d] | 43 | 740 | 240 | 391 | 640 | 125 | 318 |
| LY in 1$^{st}$ ns/Total LY | 9.2% | 60% | 31% | 49% | 22% | 2.0% | 2.5% | 1.0% | 3.3% | 1.9% | 1.3% | 1.3% |
| 40 keV Att. Leng. (1/e, mm) | 0.106 | 0.106 | 0.407 | 0.314 | 0.439 | 0.394 | 0.185 | 0.251 | 0.314 | 0.319 | 0.214 | 0.334 |

[a] top/bottom row: slow/fast component; [b] at the emission peak; [c] normalized to LYSO:Ce; [d] excited by alpha particles; [e] ceramic with 0.3 Mg at% co-doping; [f] density for composition Lu$_{0.7}$Y$_{0.3}$AlO$_3$:Ce

*zhu@caltech.edu; phone 1 626 395-6661; fax 1 (626) 395-8728; http://www.hep.caltech.edu/~zhu/

Cerium doped lutetium yttrium oxyorthosilicate ($Lu_{2(1-x)}Y_{2x}SiO_5$:Ce or LYSO) and lutetium aluminium garnet ($Lu_3Al_5O_{12}$ or LuAG:Ce) show high stopping power, high light output, fast decay time and good radiation hardness against ionization dose and hadrons. LYSO crystals are used to construct a barrel timing layer (BTL) for the CMS upgrade for the HL-LHC, where 5 Mrad ionization dose, $2.5\times10^{13}$ charged hadrons/cm$^2$ and $3\times10^{14}$ 1 MeV equivalent neutrons/cm$^2$ are expected [4]. They were also proposed for an ultra-compact, radiation hard shashlik calorimeter for the HL-LHC [5]. Yttrium doped barium fluoride crystals ($BaF_2$:Y) have an ultrafast scintillation component with 0.5 ns decay time and a suppressed slow component [6]. An ultrafast $BaF_2$:Y total absorption calorimeter is considered by the Mu2e-II experiment [3]. Cost-effective crystals and glasses are under development for the HHCAL concept. It is also interesting to note that ultrafast inorganic scintillators listed in Table 1 may also be used for GHz hard X-ray imaging at future free electron laser facilities [7].

## LYSO CRYSTALS AND LUAG CERAMICS

Radiation hardness against ionization dose, protons and neutrons are investigated for LYSO crystal and LuAG ceramic samples from various vendors. The JPL Total Ionization Dose (TID) facility, the CERN PS-IRRAD Proton Facility, and the Weapons Neutron Research facility of the Los Alamos Neutron Science Center (WNR of LANSCE) were used in this investigation. Fig. 1 shows radiation induced absorption coefficient (RIAC) values as a function of integrated ionization dose (Left), proton fluence (Middle), and 1 MeV equivalent neutron fluence (Right), for various LYSO crystal samples, and compared to the radiation hardness specification for CMS BTL LYSO crystals of 3 x 3 x 50 mm$^3$: RIAC < 3 m$^{-1}$ after 4.8 Mrad, $2.5\times10^{13}$ p/cm$^2$ and $3\times10^{14}$ n$_{eq}$/cm$^2$ [4]. In this investigation, we also found that radiation damage induced by protons in LYSO crystals is an order of magnitude larger than that from neutrons. This is due to the ionization energy loss from protons in addition to the damage caused by displacement and nuclear breakup.

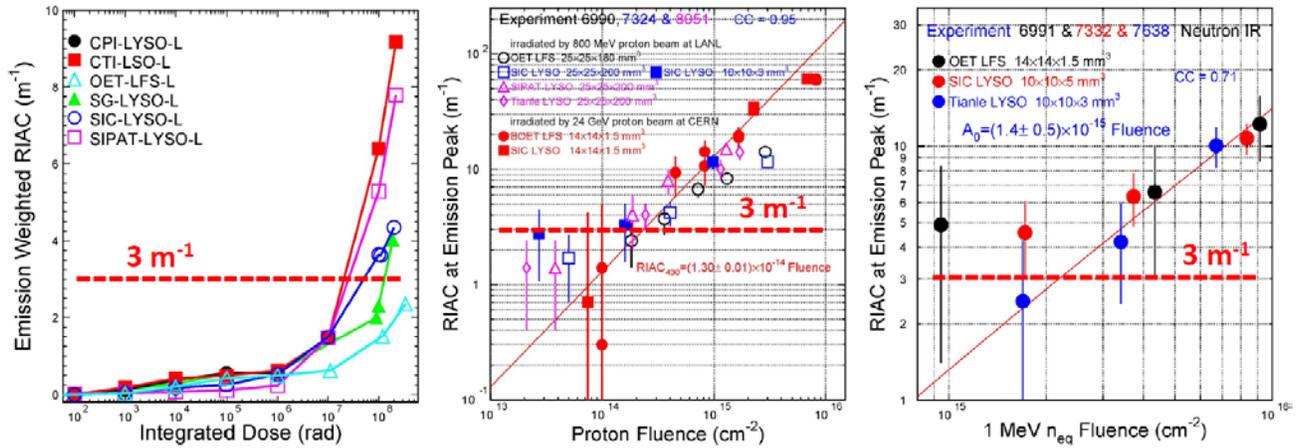

**Figure 1.** The RIAC values are shown as a function of the ionization dose (Left), proton fluence (Middle), and one MeV equivalent neutron fluence (Right) for LYSO crystals from various vendors.

Fig. 2 show RIAC values as a function of 1 MeV equivalent neutron fluence (Left) and proton fluence (Middle) up to $6.7\times10^{15}$ n$_{eq}$/cm$^2$ and $1.2\times10^{15}$ p/cm$^2$ for LuAG:Ce ceramics and LYSO:Ce and $BaF_2$ crystals. LuAG:Ce ceramics shows a factor of two better radiation hardness than LYSO crystals, so are promising for the FCC-hh [8]. LuAG:Ce ceramics, however, also has slow scintillation component. Fig. 2 (Right) shows that Ca$^{2+}$ co-doping improves the Fast/Total (F/T) ratio, defined as the ratio between the light output in 200 ns and 3,000 ns, to 90%. Investigation continues to improve the F/T ratio and radiation hardness for LuAG:Ce/Ca ceramics.

The left plot of Fig. 3 shows the LuAG:Ce excitation spectrum (red) [8] matches well with the LYSO:Ce emission spectrum (blue) [9], indicating that LuAG:Ce may serve as an effective wavelength shifter for LYSO:Ce crystals. The middle plot shows the proposed RADiCAL concept [10], which uses tungsten plates as absorber, LYSO:Ce crystals as sensitive material and LuAG:Ce ceramic fibers as wavelength shifter. The RADiCAL concept provides an ultra-compact, radiation hard and longitudinally segmented electromagnetic calorimeter for the HL-LHC and FCC-hh. The right plot shows seven $\Phi1\times40$ mm LuAG:Ce ceramic fibers under investigation, which were produced at Shanghai Institute of Ceramics (SIC) by using the laser heated pedestal growth technology for LuAG:Ce fibers.

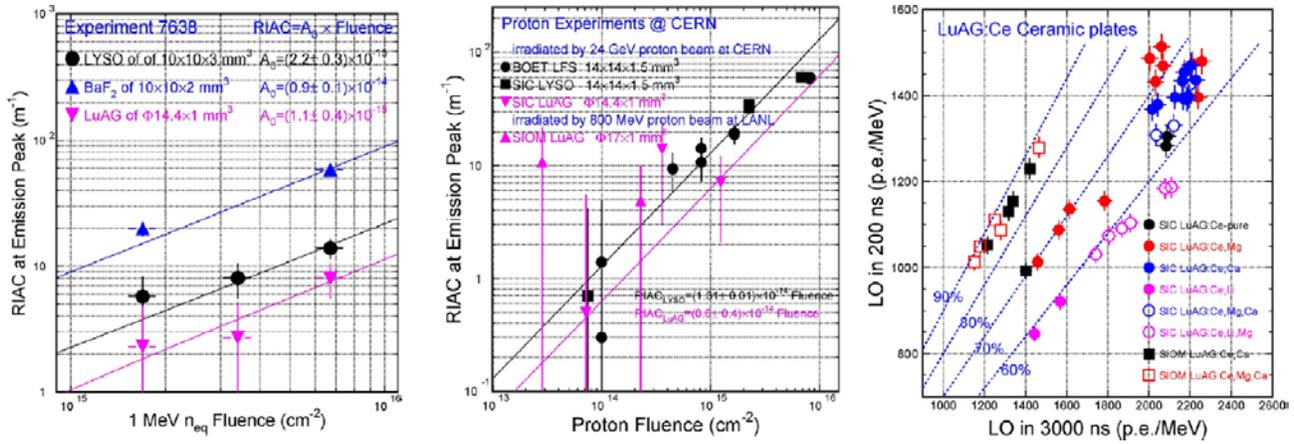

**Figure 2.** RIAC values are shown as a function of the one MeV equivalent neutron fluence (Left) and proton fluence (Middle) for LuAG:Ce, LYSO:Ce and BaF$_2$ crystals. Right: The light output in 200 ns and 3 µs gate is shown for LuAG:Ce ceramics.

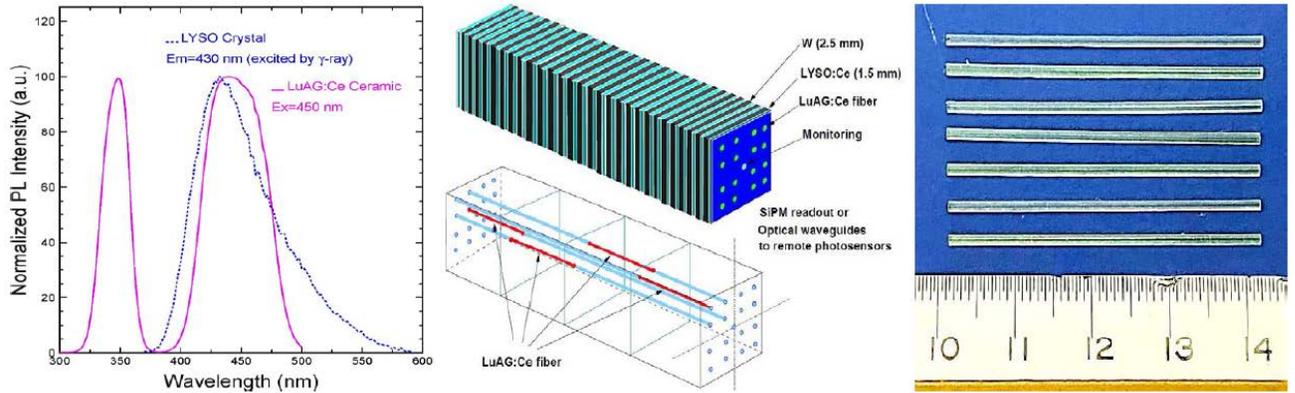

**Figure 3.** Left: The LuAG:Ce excitation (red) matches well the LYSO:Ce emission (blue). Middle: The RADiCAL concept using radiation hard LYSO:Ce as sensitive material and LuAG:Ce ceramics as wavelength shifter. Right: Seven Φ1×40 mm LuAG:Ce ceramic fibers produced at SIC by using the laser heated pedestal growth technology.

## ULTRAFAST BAF$_2$:Y CRYSTALS

It is well known that BaF$_2$ crystals have an ultrafast cross-luminescence scintillation with sub-ns decay time peaked at 220 nm, and a 600 ns slow component peaked at 300 nm with a much higher intensity. The latter causes pileup in a high-rate environment. The left plot of Fig. 4 shows the pulse shape measured by a PMT (top) and a MCP (bottom) for a BaF$_2$ sample. A decay time of 0.5 ns is observed by the MCP, but not PMT, where 1.4 ns decay time is due to the slow response time of the PMT. It is also known that the slow component in BaF$_2$ crystals may be suppressed either by rare earth doping in crystals or by using a solar blind photodetector [11]. The middle and right plots of Fig. 4 are respectively the X-ray excited emission spectra and light output as a function of integration time for BaF$_2$ cylinders of Φ18×21 mm$^3$ grown at Beijing Glass Research Institute (BGRI) with different Y$^{3+}$ doping levels [6]. They show a reduced slow light intensity for an increased yttrium doping level, while the intensity of the fast emission is maintained.

One of the potential applications of the ultrafast BaF$_2$:Y scintillation is front imager for GHz hard X-ray imaging required by future free electron laser facilities [12]. Fig. 5 show response of BaF$_2$, BaF$_2$:Y, ZnO:Ga and LYSO:Ce crystals to septuplet X-ray bunches with 2.83 ns bunch spacing measured at the advanced photon source facility of ANL. While BaF$_2$ crystals show clearly separated X-ray bunches with 2,83 ns spacing, the slow crystals do not. In addition, amplitude reduction is also observed for eight septuplets in BaF$_2$ and LYSO:Ce, which is due to MCP saturation caused by the slow scintillation in BaF$_2$ and LYSO:Ce, but not in slow-suppressed BaF$_2$:Y.

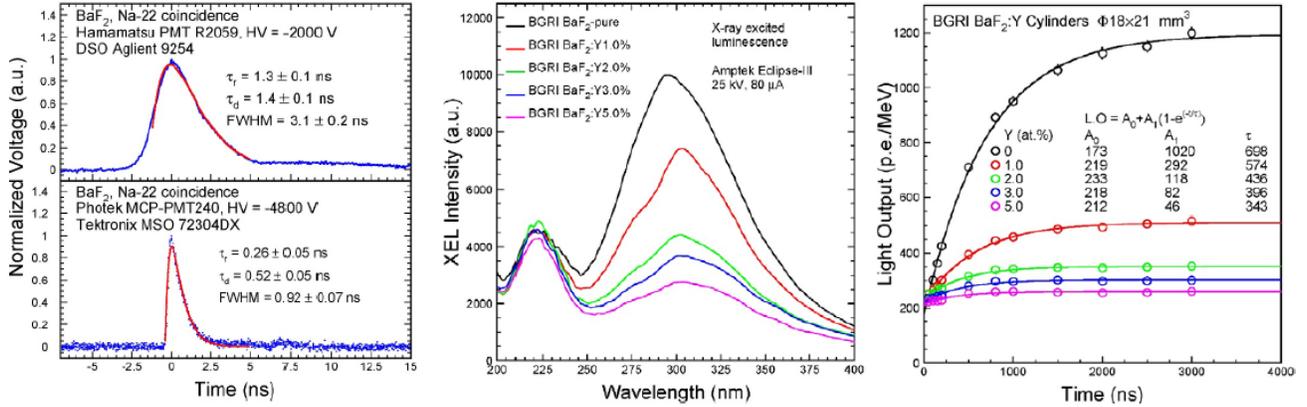

**Figure 4.** Left: The pulse shape measured by a PMT (top) and a MCP (bottom) shows the ultrafast scintillation light component with 0.5 ns decay time for a $BaF_2$ sample. The emission (Middle) and integrated light output (Right) are shown for $BaF_2$ samples with yttrium doping at different levels.

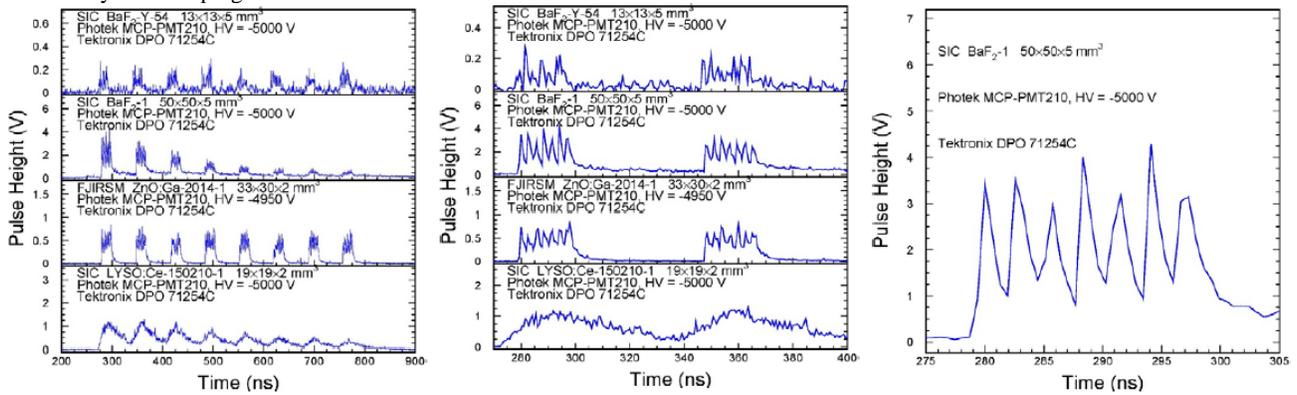

**Figure 5.** The temporal response to septuplet 30 keV X-ray bunches with 2.83 ns bunch spacing measured at APS of ANL is shown for $BaF_2$:Y, $BaF_2$, ZnO:Ga and LYSO:Ce crystal samples.

$BaF_2$:Y crystals of large size is under development for the Mu2e-II experiment [13]. In addition to yttrium doping in $BaF_2$ crystals solar-blind photodetectors also improve the F/S ratio, and thus reduce radiation induced readout noise to a level of less than 1 MeV for a $BaF_2$:Y crystal-based ultrafast calorimeter [13]. Fig. 6 shows the quantum efficiency (QE) of a Photek solar-blind cathode (Left), the particle detection efficiency (PDE) of a FBK solar-blind SiPM (Middle) and the PDE of a Hamamatsu VUV SiPM (Right) as a function of wavelength. The ability of solar-blind photodetectors in slow suppression is clearly demonstrated. R&D continues to develop solar-blind photodetectors [11].

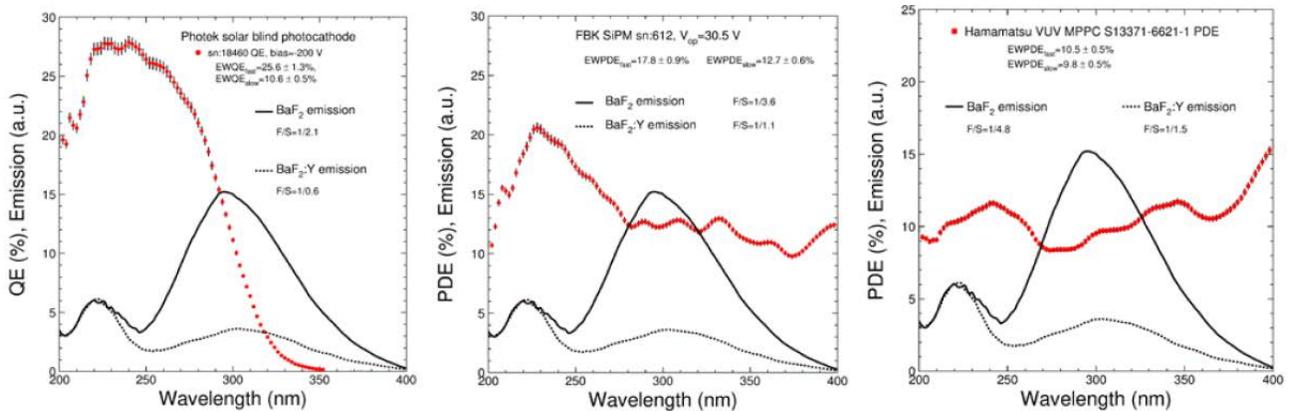

**Figure 6.** The QE of a Photek solar-blind cathode (Left), the PDE of a FBK solar-blind SiPM (Middle) and the PDE of a Hamamatsu VUV SiPM (Right) are shown as a function of wavelength.

BaF$_2$:Y crystals of large size is under development for the Mu2e-II experiment. For an ultrafast calorimetry radiation hardness against ionization dose and hadrons is crucial. The left plot of Fig. 7 shows normalized emission weighted transmittance (top) and light output (bottom) as a function of integrated ionization dose up to 130 Mrad for three 20 cm long BaF$_2$ samples from different vendors. It is interesting to note the ionization dose induced damage saturates around 10 krad until 130 Mrad, indicating limited defect density in these crystals. This is confirmed by normalized light output as a function of proton (Middle) and neutron (Right) fluence up to $9.7\times10^{14}$ p/cm$^2$ and $8.3\times10^{15}$ n$_{eq}$/cm$^2$. R&D continues to investigate radiation hardness for large BaF$_2$:Y crystals [13].

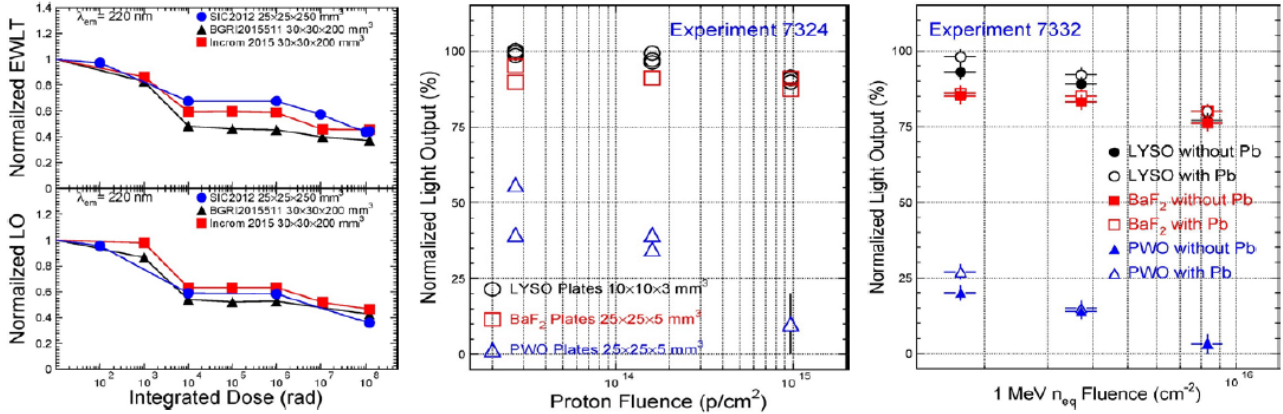

**Figure 7.** Left: Normalized transmittance (top) and light output (bottom) are shown as a function of integrated ionization dose for three 20 cm long BaF$_2$ samples from different vendors. Normalized light output is shown as a function of integrated proton (Middle) and one MeV equivalent neutron fluence (Right) for LYSO:Ce, BaF$_2$ and PWO crystal plates.

## COST-EFFECTIVE INORGANIC SCINTILLATORS

An electron-positron Higgs factory is the highest-priority next collider, where good EM and jet resolutions are required for study all decay channels of the Higgs boson. Fig. 8 shows the dual readout CalVision crystal ECAL followed by the IDEA HCAL [15], promising a 3%/√E energy resolution for electrons and photons, and a 27%/√E energy resolution for neutral pions, which is much better than 60% achieved so far by sampling hadron calorimetry.

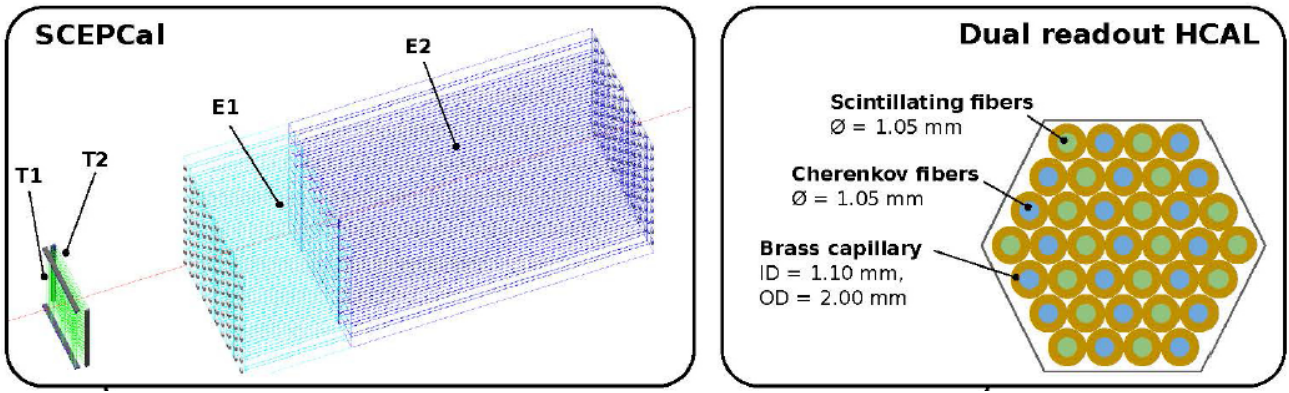

**Figure 8.** The dual readout CalVision crystal ECAL followed by the IDEA dual readout fiber HCAL.

An even better resolution for hadrons may be achieved by a total absorption hadron calorimeter. Figure 9 (Left) shows a homogeneous hadron calorimeter (HHCAL) concept [16] where a large volume of inorganic scintillators allows event by event corrections with dual readout to recover the missing energies due to nuclei breakup in hadronic jets. Extensive GEANT simulations carried out at Fermilab, ANL and CERN show that a better than 20%/√E energy resolution can be achieved for charged pions (Middle and Right) by using dual readout of scintillation and Cerenkov light, or scintillation light in short and long gates.

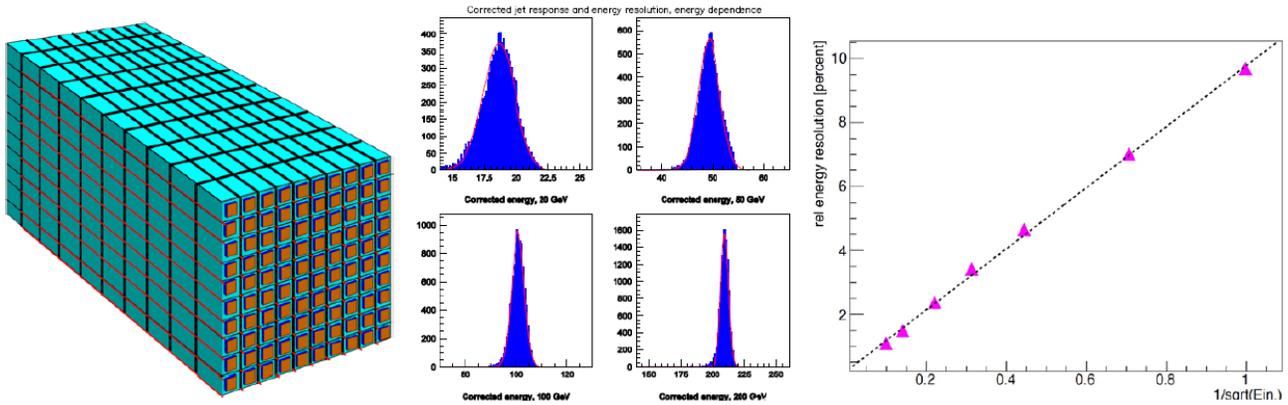

**Figure 9.** The HHCAL concept (Left) promises a 20%/√E energy resolution for charged pions (Middle) and a better than 20% stochastic term (Right) in hadronic energy resolution.

Because of the large volume of inorganic scintillators required for these total absorption calorimetry, development of cost-effective inorganic scintillators is crucial. The material of choice must be dense, cost-effective, UV transparent (for effective collection of Cerenkov light) and allow for a clear discrimination between the Cerenkov and scintillation light.

**Table 2** Optical and scintillation properties of candidate inorganic scintillators for the HHCAL concept

| | BGO | BSO | PWO | $PbF_2$ | PbFCl | Sapphire:Ti | AFO Glass | DSB:Ce Glass[1] | DSB:Ce,Gd Glass[2,3] | HFG Glass[4] |
|---|---|---|---|---|---|---|---|---|---|---|
| Density (g/cm³) | 7.13 | 6.8 | 8.3 | 7.77 | 7.11 | 3.98 | 4.6 | 3.8 | 4.7 - 5.4 | 5.95 |
| Melting point (°C) | 1050 | 1030 | 1123 | 824 | 608 | 2040 | 980[5] | 1420[6] | 1420[6] | 570 |
| $X_0$ (cm) | 1.12 | 1.15 | 0.89 | 0.94 | 1.05 | 7.02 | 2.96 | 3.36 | 2.14 | 1.74 |
| $R_M$ (cm) | 2.23 | 2.33 | 2.00 | 2.18 | 2.33 | 2.88 | 2.89 | 3.52 | 2.56 | 2.45 |
| $\lambda_I$ (cm) | 22.7 | 23.4 | 20.7 | 22.4 | 24.3 | 24.2 | 26.4 | 32.8 | 24.2 | 23.2 |
| $Z_{eff}$ value | 72.9 | 75.3 | 74.5 | 77.4 | 75.8 | 11.2 | 42.8 | 44.4 | 48.7 | 56.9 |
| dE/dX (MeV/cm) | 8.99 | 8.59 | 10.1 | 9.42 | 8.68 | 6.75 | 6.84 | 5.56 | 7.68 | 8.24 |
| Emission Peak[a] (nm) | 480 | 470 | 425 420 | \ | 420 | 300 750 | 365 | 440 460 | 440 460 | 325 |
| Refractive Index[b] | 2.15 | 2.68 | 2.20 | 1.82 | 2.15 | 1.76 | \ | \ | \ | 1.50 |
| LY (ph/MeV)[c] | 7,500 | 1,500 | 130 | \ | 150 | 7,900 | 450 | 3,150 | 2,500 | 150 |
| Decay Time[a] (ns) | 300 | 100 | 30 10 | \ | 3 | 300 3200 | 40 | 180 30 | 120, 400 50 | 25 8 |
| d(LY)/dT (%/°C)[c] | -0.9 | ? | -2.5 | \ | ? | ? | ? | -0.04 | -0.04 | -0.37 |
| Cost ($/cc) | 6.0 | 7.0 | 7.5 | 6.0 | ? | 0.6? | ? | 2.0 | 2.0? | ? |

a. Top line: slow component, bottom line: fast component.
b. At the wavelength of the emission maximum.
c. At room temperature (20°C).

1. E. Auffray, et al., J. Phys. Conf. Ser. 587, 2015
2. R. W. Novotny, et al., J. Phys. Conf. Ser. 928, 2017
3. V. Dormenev , et al., the ATTRACT Final Conference
4. E. Auffray, et al., NIMA 380 (1996),524-536
5. R. A. McCauley et al., Trans. Br. Ceram. Soc., 67. 1968
6. I. G. Oehlschlegel, Glastech. Ber. 44, 1971

Low density crystals/glasses

Table 2 lists basic optical and scintillation properties of candidate heavy inorganic material for the HHCAL concept, where BGO, BSO, PWO, $PbF_2$, PbFCl and Sapphire:Ti are crystals, AFO, DSB and HFG are scintillating glasses. The left and middle plots of Fig. 10 show a broad emission peaked at 750 nm with a decay time of 3 μs for two titanium doped Sapphire samples. With Kyropoulos growth technology, the cost of mass-produced Sapphire ingot of Φ50×55 cm is about $12,000/piece. The right photo of Fig. 10 shows mass-produced sapphire ingots in one production facility. After cutting and polishing, the price of sapphire crystals is expected to be less than $1/cc, indicating that it is possible to mass-produce inorganic scintillators cost-effectively. While its UV cut-off wavelength of 250 nm and nuclear interaction length of 24 cm are adequate for the HHCAL application, its density and radiation length of 4 g/cc and 7 cm, however, are not ideal. Work along this line will continue to search for cost-effective heavy inorganic scintillators for these novel calorimeter concepts.

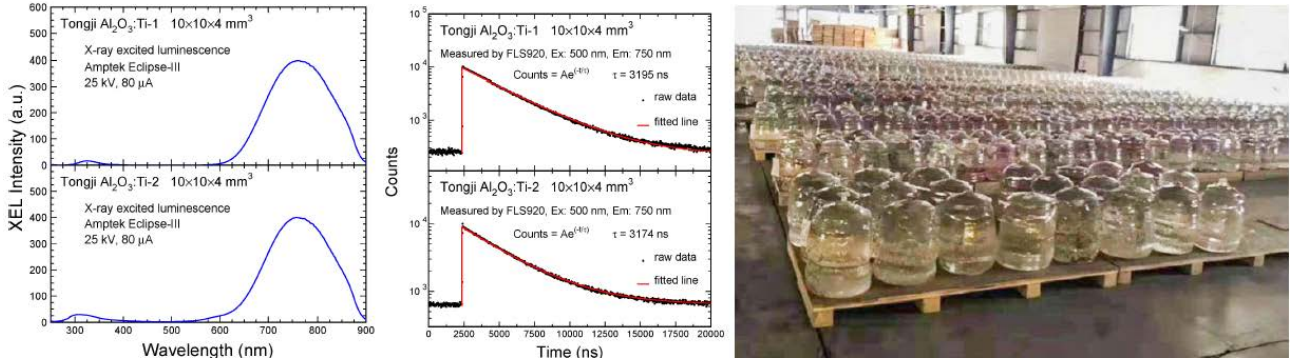

**Figure 10.** Emission (Left) and decay time (Middle) are shown for two titanium doped Sapphire crystal samples. Right: Sapphire ingots in a production facility.

## SUMMARY

Future HEP experiments at the energy frontier require fast and radiation hard calorimetry. The RADiCAL concept utilizes bright and fast LYSO:Ce crystals and LuAG:Ce WLS for an ultra-compact, ultra-radiation hard and longitudinally segmented shashlik calorimeter for HL-LHC and FCC-hh.

Future HEP experiments at the intensity frontier requires ultrafast calorimetry. R&D is on-going to develop large size $BaF_2$:Y crystals and solar-blind VUV photodetectors for Mu2e-II.

The proposed lepton Higgs factory requires good EM and jet resolutions. The dual readout CalVision crystal ECAL followed by the IDEA HCAL provides an excellent option.

Because of total absorption for hadrons the HHCAL concept promises the best jet mass resolution. Crucial R&D is to develop cost-effective inorganic scintillators of large volume.

Novel inorganic scintillators are needed for all these calorimeter concept.

## ACKNOLEDGEMENTS

This work is supported by the U.S. Department of Energy, Office of High Energy Physics program under Award Number DE-SC0011925.